\documentclass[journal]{IEEEtran}
\newcommand{\eg}{\emph{e.g.}, }

\usepackage[utf8]{inputenc}
\DeclareUnicodeCharacter{200E}{}

\ifCLASSINFOpdf
   \usepackage[pdftex]{graphicx}
   \graphicspath{{../pdf/}{../jpeg/}}
   \DeclareGraphicsExtensions{.pdf,.jpeg,.png}
\else
\fi
\usepackage{amsmath}
\usepackage{multirow}
\usepackage{amsfonts}
\usepackage{cite}

\usepackage{algorithmic}

\usepackage{array}

\ifCLASSOPTIONcompsoc
 \usepackage[caption=false,font=normalsize,labelfont=sf,textfont=sf]{subfig}
\else
 \usepackage[caption=false,font=footnotesize]{subfig}
\fi
\usepackage{fixltx2e}
\usepackage{url}

\begin{document}
%
\title{Lumbar Bone Mineral Density Estimation from Chest X-ray Images: Anatomy-aware Attentive Multi-ROI Modeling}
%
%
%

 
\author{Fakai Wang*, Kang Zheng, Le Lu, \IEEEmembership{Fellow, IEEE}, Jing Xiao, Min Wu, \IEEEmembership{Fellow, IEEE}, \\ Chang-Fu Kuo* and Shun Miao

\thanks{Fakai Wang and Min Wu are with ECE Department, University of Maryland, College Park, MD 20742, USA. (email:jackwangumd@gmail.com). Kang Zheng is with PAII Inc., Bethesda, MD 20817, USA. Le Lu is with Alibaba Group DAMO Academy, New York, NY 10014, USA. Shun Miao is with Xpeng Motors, San Diego, CA, USA.} 
\thanks{Jing Xiao is with Ping An Technology, Shenzhen, 510852, China.}
\thanks{Chang-Fu Kuo is with Chang Gung Memorial Hospital, Linkou, Taiwan, ROC. (email:zandis@gmail.com)}
\thanks{This work was partially done when Fakai Wang was an intern at PAII Inc. Chang-Fu Kuo and  Fakai Wang are the corresponding authors.}
}

\maketitle

\begin{abstract}
Osteoporosis is a common chronic metabolic bone disease often under-diagnosed and under-treated due to the limited access to bone mineral density (BMD) examinations, \eg via Dual-energy X-ray Absorptiometry (DXA). This paper proposes a method to predict BMD from Chest X-ray (CXR), one of the most commonly accessible and low-cost medical imaging examinations. Our method first automatically detects Regions of Interest (ROIs) of local CXR bone structures. Then a multi-ROI deep model with transformer encoder is developed to exploit both local and global information in the chest X-ray image for accurate BMD estimation. Our method is evaluated on 13719 CXR patient cases with ground truth BMD measured by the gold standard DXA. The model predicted BMD has a strong correlation with the ground truth (Pearson correlation coefficient \textit{0.894} on lumbar 1). When applied in osteoporosis screening, it achieves a high classification performance (average AUC of \textit{0.968}). As the first effort of using CXR scans to predict the BMD, the proposed algorithm holds strong potential for early osteoporosis screening and public health promotion.

\end{abstract}


%
\IEEEpeerreviewmaketitle

\section{Introduction}
\IEEEPARstart{O}{steoporosis} is the most common chronic metabolic bone disease, characterized by low bone mineral density (BMD) and decreased bone strength. With an aging population and longer life span, osteoporosis is becoming a global epidemic, affecting more than 200 million people worldwide~\cite{sozen2017overview}. Osteoporosis increases the risk of fragility fractures, which are associated with disability, fatality, reduced life quality, and financial burden to the family and the society. While with an early diagnosis and treatment, osteoporosis can be prevented or managed, osteoporosis is often under-diagnosed and under-treated among the population at risk~\cite{lewiecki2019challenges}. More than half of insufficiency fractures occur in individuals who have never been screened for osteoporosis ~\cite{smith2019screening}. The under-diagnosis and under-treatment of osteoporosis are mainly due to 1) low osteoporosis awareness and 2) limited accessibility of Dual-energy X-ray Absorptiometry (DXA) examination. 

Opportunistic screening of osteoporosis is an emerging research field in recent years~\cite{cheng2020opportunistic,dagan2020automated,jang2019opportunistic,pickhardt2020automated}. It aims at reusing medical images originally taken for other indications to screen for osteoporosis, which offers an opportunity to increase the screening rate at no additional cost. As the most commonly prescribed medical image scanning, plain films' excellent spatial resolution permits the delineation of fine bony micro-structure that may correlate well with the BMD. We hypothesize that specific regions of interest (ROI) in the standard chest X-rays (CXR) may help the osteoporosis screening. 

This work introduces a method to estimate the BMD from CXR to screen osteoporosis. Our method first locates anatomical bone landmarks and extracts multiple ROIs as imaging biomarkers for osteoporosis. Then We propose a novel network architecture that jointly processes the ROIs with learnable feature weight adjustment to estimate the BMDs. We experiment on 13719 CXRs with paired DXA BMDs (ground truth). This paper extends from a preliminary work~\cite{fakai2021chestBMD}. In summary, our contributions are three-fold: 1) to our best knowledge, we are the first to develop models using CXR to estimate BMD 2) we propose the anatomy-aware Attentive Multi-ROI model to combine global and local information for accurate BMD estimation. 3) Our method achieves clinically useful osteoporosis screening performance.

\begin{figure*}[bt!]
	\centering
	\includegraphics[width=0.95\linewidth]{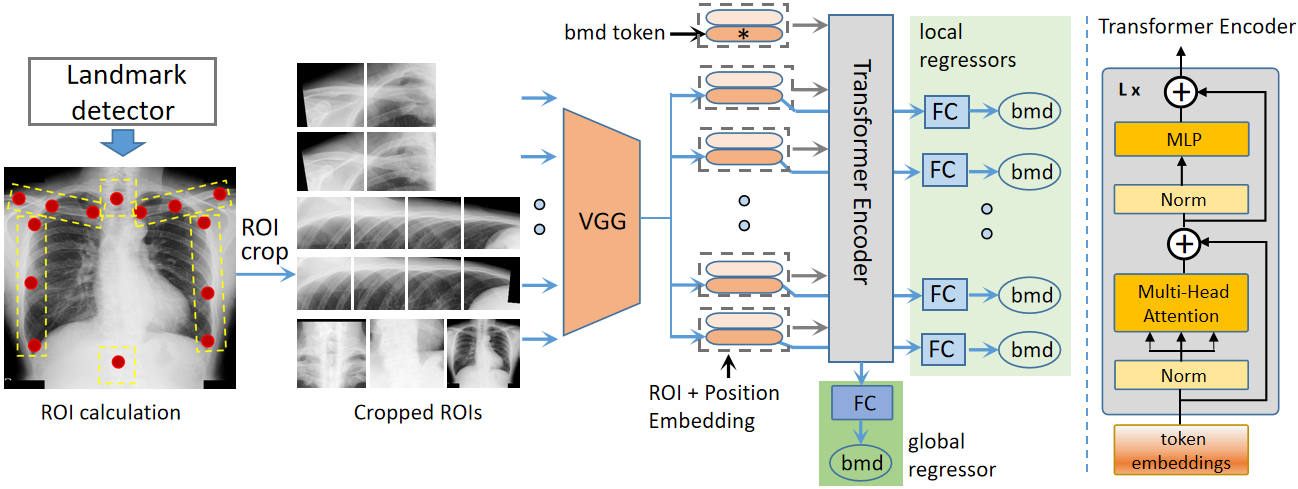}
	\caption{The proposed working pipeline. The landmark detector locates key bone points (red dots) on CXR images, then we crop and normalize 14 ROIs. These modalities go through a shared feature extractor (VGG16). The global regressor (dark green) works on global feature generated by the Transformer Encoder (grey). During training, results of both local regressors (light green) and the global regressor are used for loss calculation and back propagation.}
	\label{fig:system}
\end{figure*}

\section{Related work}
\subsection{Bone Mineral Density estimation and early screening}
BMD examination via DXA machines is essential for osteoporosis determination and fracture risk assessment. In practice, bone densities of young adults are used as the reference where the standard deviations (SD) are used as the measuring unit (T-score)~\cite{who1994assessment}. \textit{1} unit of the T-score represents \textit{1} unit of SD of the density, and \textit{0} T-score represents the mean density of all the references. T-score at the spine, hip ,or mid-radius lower than \textit{-2.5} is considered osteoporosis. T-score between -2.5 and -1 is considered osteopenia, and T-score above -1 is considered normal. However, when used for general screening, DXA BMD examination has many shortcomings. First and foremost, DXA services are not widely available for general screening. Second, the mere DXA-based diagnosis could not ensure accurate in-time bone quality evaluation, since more fractures occur without reaching the severity of the osteoporosis~\cite{Compston2017UKCG}\cite{smith2019screening}, due to the complex causes and symptoms. 

Many works have discussed osteoporosis screening from non-DXA examinations. Firstly, the Quantitative Computed Tomography (QCT) in abdomen or chest can be re-used without additional radiation exposure or cost~\cite{Boutroy2005InVA}\cite{Schreiber2011HounsfieldUF}\cite{Adams2009QuantitativeCT}\cite{Lee2013CorrelationBB}\cite{Li2018OpportunisticSF}. The \textit{Hounsfield Units} of the QCT scans correlate well with DXA BMD scores for low BMD diagnosis~\cite{Schreiber2011HounsfieldUF}\cite{Alacreu2016OpportunisticSF}. QCT has the advantage of the three-dimensional assessment of the structural and geometric properties of the examined bone~\cite{Specker2005QuantitativeBA}. Secondly, the Quantitative Ultra Sound (QUS) based techniques have advantages of safety, low cost, operating flexibility~\cite{Specker2005QuantitativeBA}\cite{NatureBaroncelli2008Quantitative}\cite{Pisani2013ScreeningAE}. X-ray based examinations (DXA, QCT) are not suitable for the sensitive such as young children or the pregnant because of ionizing radiation. DXA or CT machines occupy extensive space and require specially trained operators, impeding general screening. QUS avoids these shortcomings, but QUS methods do not have standards on skeletal measuring sites, performance criteria, or normative reference data in the clinical setting~\cite{NatureBaroncelli2008Quantitative}. 

Lastly, the plain film or X-ray, as the most common radiography examination, can also be utilized for osteoporosis screening. The existing techniques (DXA, QCT, QUS) all work on specific bone regions concerning the score from areal or volumetric mass. The X-ray images however contain not only the bone textures but also other contexts. Since osteoporosis is a metabolic bone disease with complex manifestation, involving a larger context could capture the density relation which benefits BMD estimation. Hip X-ray based BMD estimation have been verified in osteoporosis screening~\cite{kang2021semiBMDhip}. In this paper we investigate the chest X-ray based BMD estimation with a focus on the input modality, model architectures, and prediction applicability.

\subsection{Convolutional neural network and self-attention mechanism}
Convolutional Neural Networks (CNN) have succeeded in medical image analysis~\cite{Ronneberger2015UNet}\cite{Shin2015textImageDatabase}\cite{Yan2018DeepLesionAM}\cite{Yamashita2018ConvolutionalNN}, partly because the hierarchical visual patterns echo the inductive biases learned by CNN layers. However, the inductive biases including translation equivalence and locality are less important for BMD pattern learning. The texture contrast among neighboring pixels and regions has more BMD cues. But the bare CNN backbones operate locally, failing to compare regional contexts ~\cite{xiaolongwang2018NonLocalNet}\cite{Bello2019ICCVAttentionAugmentedCNN}\cite{Vit}. Some papers exploit textual relationship by enhancing spatial feature encodings~\cite{xiaolongwang2018NonLocalNet} or through channel-wise feature recalibration~\cite{Hu2018SENet}. CCNet~\cite{Huang_2019_ICCV_CCNet} proposes the criss-cross attention module to harvest the contextual information on the criss-cross path. LR-Net~\cite{Hu2019LocalRN} presents the local relation layer (Local Relation Network) that adaptively determines aggregation weights based on the compositional relationships. GCNet~\cite{CaoYue2019GCNet} unifies the simplified non-local network~\cite{xiaolongwang2018NonLocalNet} and SENet~\cite{Hu2018SENet} into a general framework for global context modeling. 

Inspired by the \textit{Transformer} success in language tasks~\cite{transformer}\cite{Bert}\cite{Radford2019LanguageGPT2}\cite{Brown2020LanguageMAGPT3}, emerging works employ the \textit{Transformer} modules to replace or facilitate the convolutional layers for visual tasks~\cite{Vit}\cite{Hu2017RelationNetForObjDet}\cite{Han2020ASOViT}\cite{khan2021transformers}. The \textit{Transformer Encoder} learns the global relationship through repetitive layers of \textit{Multi-Head Self-attention} and \textit{Multi-Layer Perception} operations. Attention Augmentation~\cite{Bello2019ICCVAttentionAugmentedCNN} augments convolutional operators with self-attention mechanism by concatenating convolutional feature maps with a set of self-attention feature maps. The iGPT~\cite{Chen2020iGPT} train the transformer model~\cite{Radford2019LanguageGPT2} on pixel sequences to generate coherent image completions in unsupervised settings. 

The CXR BMD task requires the model to capture both local textures and regional relations automatically. A mindfully tailored combination of CNN and Transformer could harness their strengths. Convolutional layers can capture inductive biases such as translation equivariance and locality, while the transformer encoder enables global feature interaction. Therefore, we employ both the convolutional feature extractor and self-attention fusion module in our proposed \textit{Attentive Multi-ROI} model.

\section{Methodology}
\subsection{Task Overview}
\begin{figure}[bt!]
	\centering
	\includegraphics[width=0.7\linewidth]{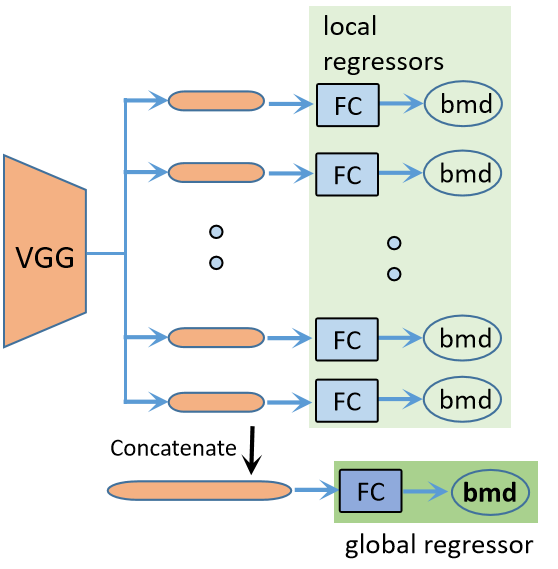}
	\caption{The plain fusion process in the Multi-ROI model. Individual feature vectors are concatenated as one before going through the global regressor.}
	\label{fig:plain_fusion}
\end{figure}

\begin{figure}[bt!]
	\centering
	\includegraphics[width=0.80\linewidth]{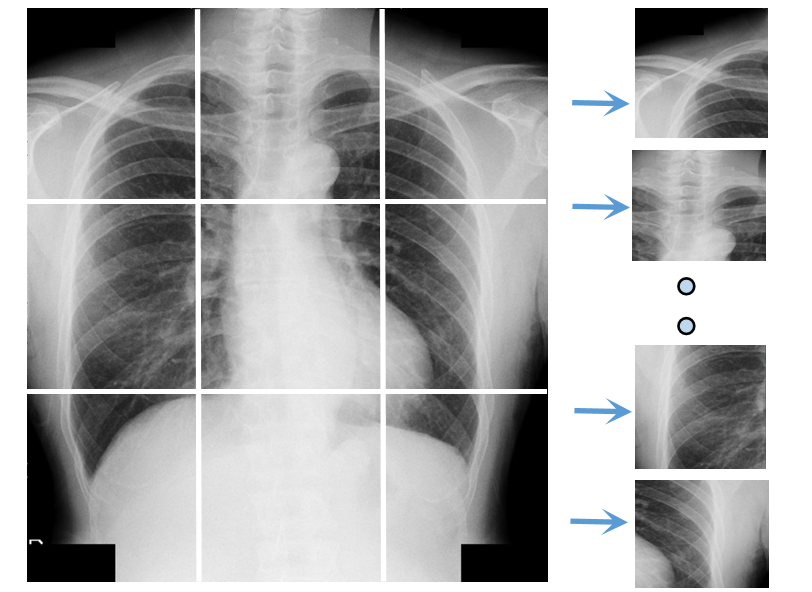}
	\caption{Patch generation of Multi-Patch model. The CXR image is split into \textit{3}x\textit{3} non-overlapping regions.}
	\label{fig:image_patches}
\end{figure}

In the opportunistic screening setting, the input is a chest X-Ray image. Our goal is to predict the BMD of lumbar vertebrae (L1, L2, L3, L4), alarming the patient of possible low BMD or osteoporosis conditions. Our hypothesis is that the BMD information lies in the CXR patterns, both in individual bone textures (local information) and in the overall combination of chest bone contexts (global information). Directly feeding the whole chest image into a CNN model for BMD prediction (\textbf{Baseline}) is intuitive and viable, but it lacks localized and contrastive bone information among the chest regions. The proposed pipeline in Figure~\ref{fig:system} consists of chest landmark detection, bone ROI cropping, local texture extraction, global feature fusion, BMD regression. We get the key landmarks for the chest from a Graph Convolution Network (GCN) model~\cite{li2020structured}. Then bone regions are cropped accordingly and fed into the proposed Attentive Multi-ROI model. The CNN layers learn local textures and patterns, extracting individual features. The Transformer Encoder refines individual representations to enable inter-regional feature interaction. At the end of the pipeline, local regressors and global regressor make BMD predictions on corresponding features. The model variants (in Fig~\ref{fig:plain_fusion} Fig\ref{fig:image_patches}) are studied in Experiments section.

\subsection{Automatic ROI Extraction} 
\label{ssec:implement_landmarkandROI}
As the first step in the proposed pipeline, bone selection and region cropping prepare the model inputs. There are multiple bones in the chest area, bearing varied importance for BMD prediction. Although all bones provide density information due to the metabolic nature, the model should focus on the most effective regions. It is also unclear if the combination of distinct bone patterns are essential for this task. To learn representations of the local textures and to explore the correlation among different regions, medical experts advise us to extract ROIs for clavicle bone, cervical vertebra, lumbar vertebra, ribcage edges. We avoid the central part of the chest X-ray where cardiac or pulmonary diseases may significantly influence the appearance. In the end, our model works on the ROI croppings of left/right clavicle bones, cervical spine, left/right rib-cage area, T12 vertebra. 

We employ the Graph Convolution Network (GCN) based Deep Adaptive Graph (DAG)~\cite{li2020structured} to automatically detect critical landmarks in the chest. We identify 16 landmarks in Figure~\ref{fig:system}, which include 1) 3 points on the left/right clavicles, 2) 4 points along the left/right rib cages, 3) 1 point on the C7 vertebrae, 4) 1 point on the T12 vertebra. We manually labeled 1000 cases (16 landmarks on each CXR scan) as the training samples for the DAG model. The resulting landmark detector can reliably extract all the keypoints. Given the keypoints for each bone, we crop the corresponding bone regions. However, different bones have distinct shapes and sizes, so we further sub-split the wider and higher ones. As seen in the cropped ROIs~\ref{fig:system}, there are 2, 2, 4, 4, 1, 1 croppings for left clavicle, right clavicle, left ribcage, right ribcage, cervical, lumbar respectively. This arrangements are based on the bone size and width/height ratio. Besides these 14 local ROIs, we also include the whole CXR image as one modality. These 15 ROIs are resized and normalized before going through CNN layers. 

\subsection{ Hybrid architecture of convolution and self-attention} 
The feature extractor backbone is VGG16, and we extract local patterns separately for each bone region (ROI). Since there is little variation in the chest outlines among different people, the overall appearance alone is not enough to determine BMD. Instead, finer level texture and pixel densities around the bones tell the distinctions between the normal and the osteoporosis. So individual ROI features are extracted independently. VGG16 is better than more complex backbones because its relatively shallow layers fit the simplicity of bone texture characteristics. We use \textit{average\_pool} to reduce the spatial dimensions on the VGG16 feature outputs, generating local representations $\textbf{f}_i,\ i \in (1,..,N)$, \textit{N}=15. However, the local textures and pattern combinations could have distinct manifestations in different people for the same BMD value~\cite{who1994assessment}\cite{Compston2017UKCG}. To make reliable density predictions, it should be addressed from the global level to account for the intractable variations resulting from diseases, scanning settings, and noises. 

We employ the \textit{Transformer Encoder} (grey box in Figure~\ref{fig:system}) in the global feature integration. The feature fusion process adjusts the individual features through layers of Multi-head Self-Attention (MSA) and Multiple Linear Perception (MLP) units, where the weighted relations are learned automatically. Similar to \textit{BERT}'s \textit{[class token]}, we prepend a learnable \textit{bmd token} embedding $\textbf{E}_{bmd}$ as the target holder to increase robustness. In Equation~\ref{eqn:encoderembedding}, $\textbf{f}_i$ is the feature representation for the $i$th ROI, $\textbf{E}_{bmd}$ is the learnable \textit{bmd token} (\textit{target holder}, orange box with * inside in Figure~\ref{fig:system}), $\textbf{E}_{pos}$ represent $N$=16 learnable positional embeddings, $\textbf{z}_0$ is the initial embeddings fed into the transformer encoder. The learnable position embeddings $\textbf{E}_{pos}$ are essential to keep spatial identity during the self-attention computation since there is no explicit sequential or grammatical order in the visual patches. In Equation~\ref{eqn:encoderMSA} and \ref{eqn:encoderMLP}, the alternating operations of MSA and MLP refine the feature representations. In our proposed model the encoder consists of \textit{L}=6 layers similar to~\cite{Vit}. And each layer consists of \textit{Layer Norm}, \textit{MSA}, \textit{Layer Norm}, \textit{MLP}. In Equation~\ref{eqn:encoderFeatGlobal}, the \textit{mean} of the adjusted feature embeddings are used as the global feature representation.  

\begin{equation}
\begin{split}
\textbf{z}_0 = [\textbf{E}_{bmd}; \textbf{f}_1; \textbf{f}_2; ...; \textbf{f}_N] + \textbf{E}_{pos}
\label{eqn:encoderembedding}
\end{split}
\end{equation}
\begin{equation}
\begin{split}
\textbf{z}'_l = MSA(LN(\textbf{z}_{l-1})) + \textbf{z}_{l-1}, \ \ \ l = 1 ... L 
\label{eqn:encoderMSA}
\end{split}
\end{equation}
\begin{equation}
\begin{split}
\textbf{z}_l = MLP(LN(\textbf{z}'_l)) + \textbf{z}'_l, \ \ \ l = 1 ... L 
\label{eqn:encoderMLP}
\end{split}
\end{equation}

\begin{equation}
\begin{split}
\textbf{f}_{global} = \frac{1}{N} \sum_{i \in 1,...N} \textbf{z}_L^{i} 
\label{eqn:encoderFeatGlobal}
\end{split}
\end{equation}

The core module of the transformer encoder is the \textit{Multi-head Self-Attention}, illustrated in the \textit{QKV} form in Equation~\ref{eqn:qkv}-\ref{eqn:msa}. In Equation~\ref{eqn:qkv}, $\textbf{z}\in \mathbb{R}^{N\times D}$ is the feature embedding, the \textit{query (\textbf{q})}, \textit{key (\textbf{k})}, \textit{value (\textbf{v})} are the projection of embedding \textbf{z} on matrix mapping $\textbf{U}_{qkv}$. $k=6$ is the \textit{head} number, $D_h = D/k$ is the feature size in each head. In Equation~\ref{eqn:softmax}, $A$ is the weight matrix and ${A}_{ij}$ represents the pairwise similarity between the $i$th and the $j$th features. In Equation~\ref{eqn:softmax} and \ref{eqn:sa}, the Self-Attention (SA) module adjusts feature embedding according to weighted affinity. Regions of higher correlation contribute more, and others share less feature exchange. In this way, the individual feature becomes more robust to patient variations or noises. In the transformer encoder layer, the MLP consists of two linear layers with one non-linear \textit{GELU} layer. The Transformer Encoder modules uses constant latent vector size D=512 through all layers, the same size as VGG16 feature. The Layer Norm (LN) and residual connections are applied before and after every block respectively, to keep the training stable.

\begin{figure*}[bt!]
	\centering
	\hfill
	\includegraphics[width=0.32\linewidth]{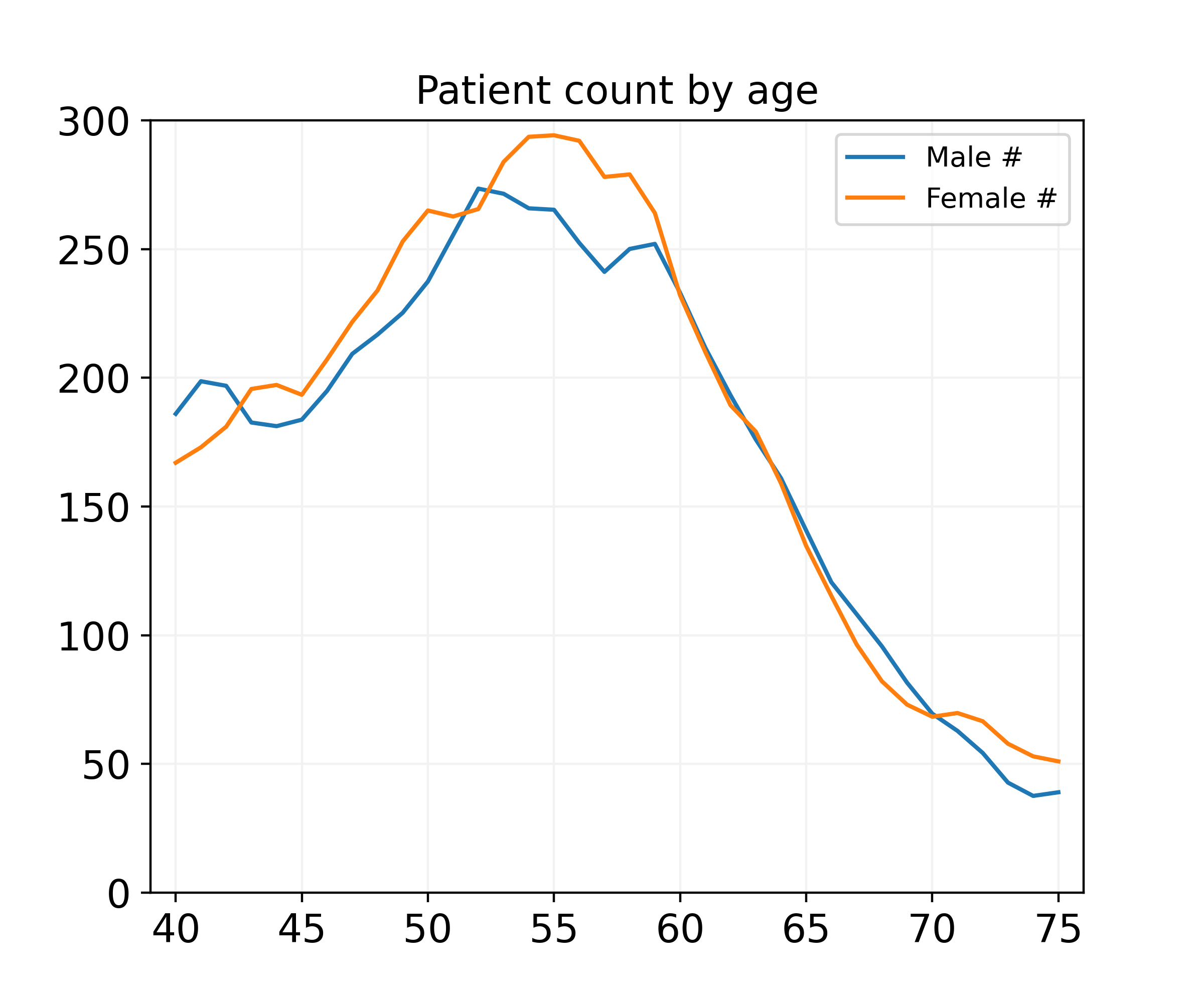}
	\hfill
	\includegraphics[width=0.32\linewidth]{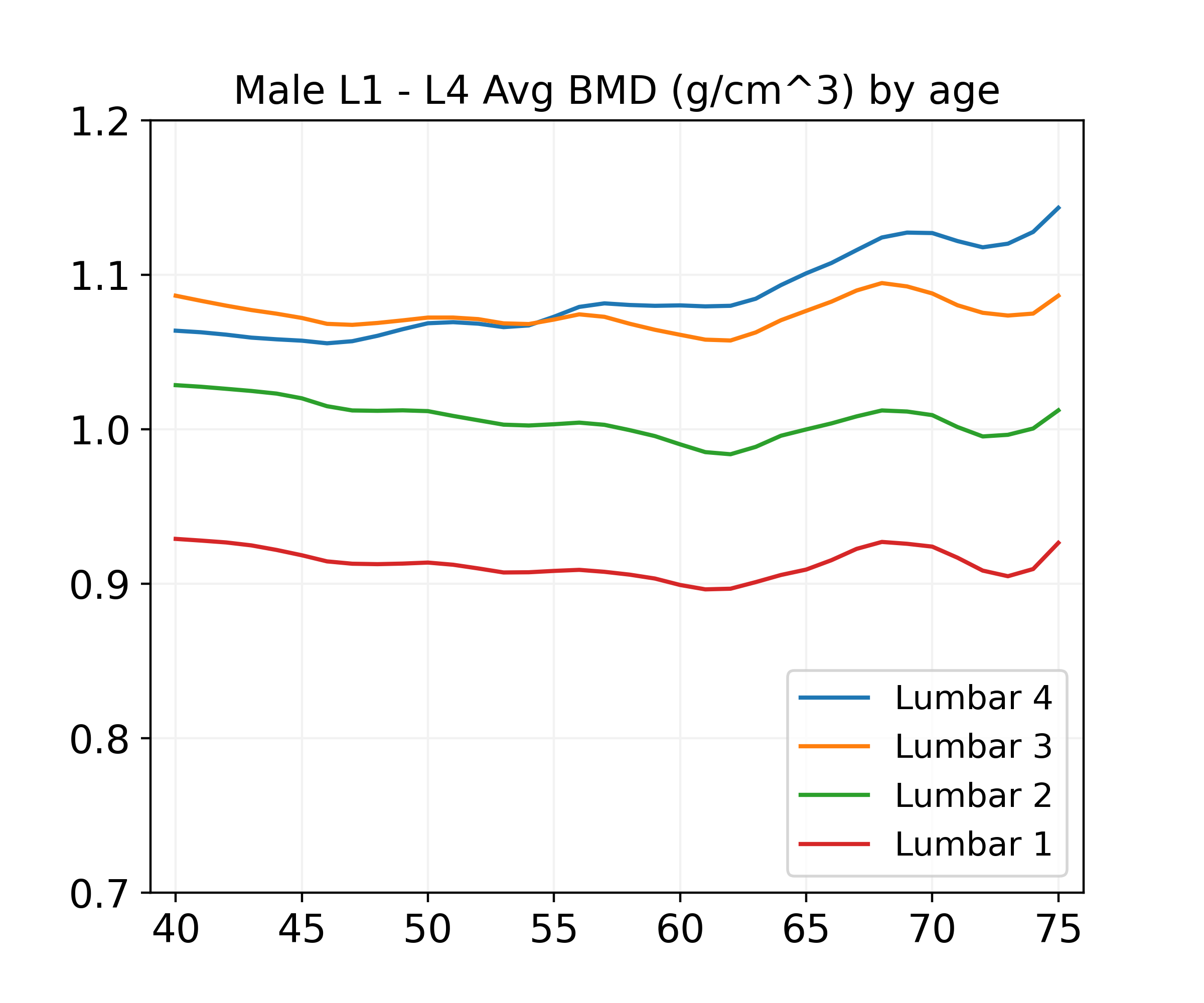}
	\hfill
	\includegraphics[width=0.32\linewidth]{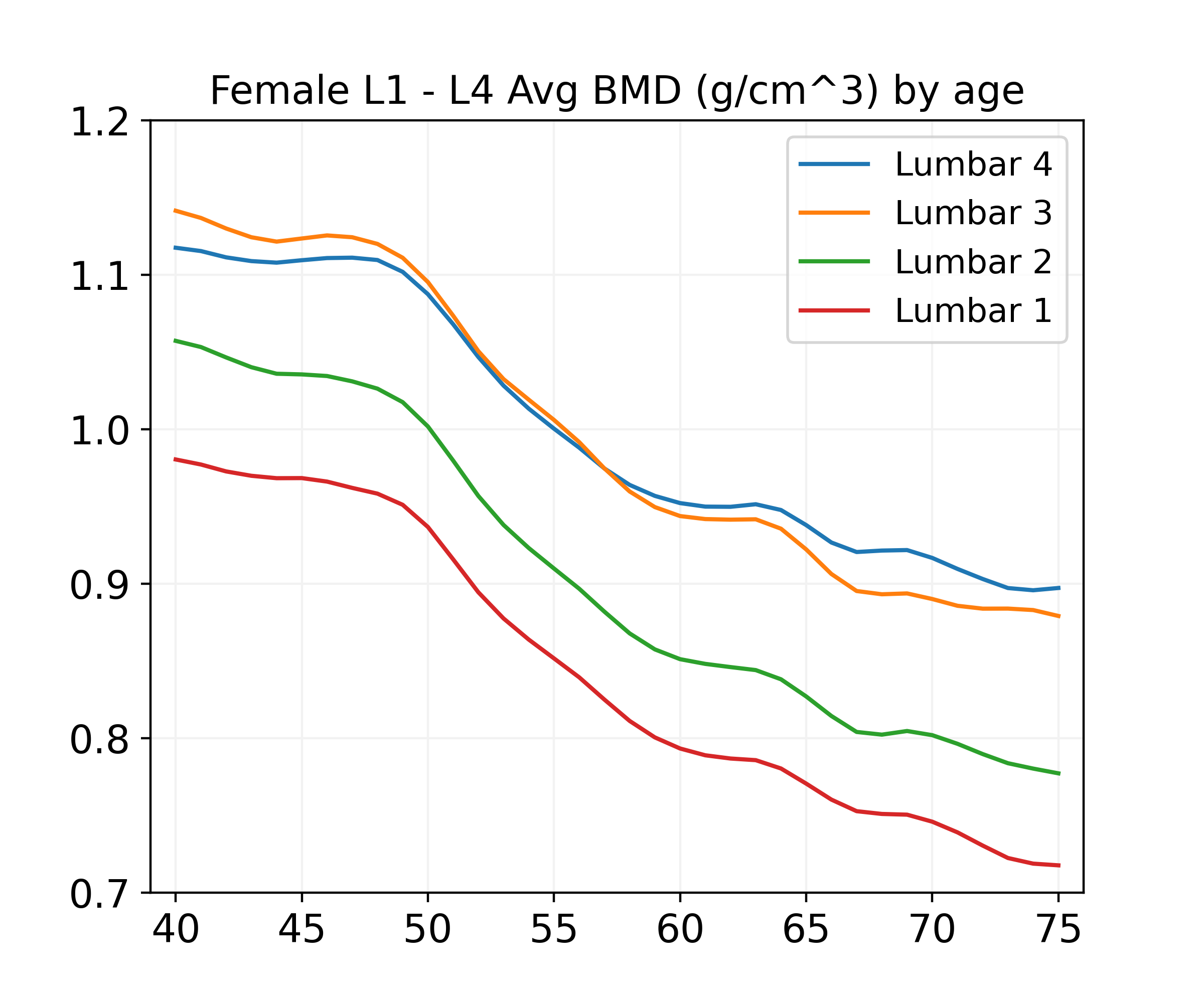}
	\hfill
	\caption{DXA BMD averages across ages for both genders. Different lumbar vertebrae (L1, L2, L3, L4) are drawn separately.}
	\label{fig:data_age_bmds}
\end{figure*}

\begin{equation}
\begin{split}
[\textbf{q,k,v}] = \textbf{z} \textbf{U}_{qkv} \ \ \ \ \  \textbf{U}_{qkv} \in \mathbb{R}^{D\times 3D_h} 
\label{eqn:qkv}
\end{split}
\end{equation}

\begin{equation}
A = softmax(\textbf{q}\textbf{k}^T / \sqrt{D_h}) \ \ \ \ A \in \mathbb{R} ^{N \times N} 
\label{eqn:softmax}
\end{equation}

\begin{equation}
SA(\textbf{z}) = A \textbf{v} 
\label{eqn:sa}
\end{equation}

\begin{equation}
    MSA(\textbf{z}) = [SA_1(\textbf{z});SA_2(\textbf{z});...;SA_k(\textbf{z})]\textbf{U}_{msa} 
    \label{eqn:msa}
\end{equation}
\ \ \ \ \ \ \ \ \ \ \ \ \ \ \ \ \ \ \ \ \ \ \ \ \ \ \ \ \ \ \ \ \ \ \ \ \ \ \ \ \ \ \ $\textbf{U}_{msa} \in \mathbb{R} ^{k\cdot D_h \times D}$

\subsection{BMD Estimation via Joint Analysis of the ROIs}
\label{ssec:multiROI_model}
We apply local regressors and global regressor on the local features $\textbf{f}_i, i \in 1,..,N$ and aggregated global feature $\textbf{f}_{global}$ respectively for BMD prediction. In Figure~\ref{fig:system}, we employ separate \textit{FC} regressors for different local ROIs. All the regressors consist of two linear layers and one \textit{ReLu} non-linear layer. We employ L2 loss on the predictions. The local regressions are active during training iterations to regularize feature representations but are ignored during evaluation, while the global regression is used all the time. During validation or inference, only the global BMD output is used. By jointly utilizing the global and local ROIs in the chest and jointly promoting feature exchange through the transformer encoder, the network is capable of extracting BMD patterns on different scales for robust regression. 

\subsection{Implementation Details}
We work on a workstation with Intel Xeon W-2295 CPU @ 3.00GHz, 132 GB RAM, and 4 NVIDIA Quadro RTX 8000 GPUs. Our models are implemented with PyTorch. The input images/ROIs sizes are set as \textit{(256, 256) } by default for the best results. The training augmentations include scaling, rotation, translation, and random flip. The SGD optimizer has a learning rate of $0.0001$, a weight decay of 4e-4. All models are trained for $100$ epochs. The four components in our model, VGG16 feature extractor, transformer encoder, local regressors, global regressors, occupy 14.8M, 7.9M, 2.1M, and 0.13M parameters respectively, which sum to 25M parameters.

\begin{table*}[h]
	\caption{Performance comparison of different models. VGG16 is the feature extractor. Our proposed (the Attentive Multi-ROI model, \textbf{Proposed}) outperforms others in all four lumbar BMD estimation task, in terms of R-value, RMSE, AUC score. }
	\centering
	\scriptsize
\begin{tabular}{|c|c|c|c|c|c|c|c|c|c|c|c|c|c|c|c|}
\hline
\multirow{2}{*}{Model} & \multicolumn{3}{c|}{L1} & \multicolumn{3}{c|}{L2} & \multicolumn{3}{c|}{L3} & \multicolumn{3}{c|}{L4} & \multicolumn{3}{c|}{Average} \\\cline{2-16} 
                       & R-Val  & RMSE  & AUC   & R-Val  & RMSE  & AUC   & R-Val  & RMSE  & AUC   & R-Val  & RMSE  & AUC   & R-Val   & RMSE    & AUC     \\\hline
Base                   & 0.859  & 0.089 & 0.952 & 0.87   & 0.101 & 0.96  & 0.86   & 0.104 & 0.966 & 0.823  & 0.12  & 0.963 & 0.853   & 0.104   & 0.96    \\\hline
MultiPatch             & 0.873  & 0.071 & 0.958 & 0.877  & 0.08  & 0.967 & 0.873  & 0.083 & 0.971 & 0.834  & 0.098 & 0.965 & 0.864   & 0.083   & 0.965   \\\hline
AttMultiPatch          & 0.885          & 0.068          & 0.964          & 0.888          & 0.076          & \textbf{0.967} & 0.882          & 0.081          & 0.969          & 0.846          & 0.094          & 0.964          & 0.875          & 0.08           & 0.966          \\\hline
MultiROI               & 0.883  & 0.068 & 0.959 & 0.887  & 0.076 & 0.966 & 0.879  & 0.082 & 0.972 & 0.837  & 0.097 & 0.966 & 0.871   & 0.081   & 0.966   \\\hline
\textbf{Proposed}      & \textbf{0.894} & \textbf{0.065} & \textbf{0.964} & \textbf{0.899} & \textbf{0.073} & 0.966          & \textbf{0.887} & \textbf{0.079} & \textbf{0.972} & \textbf{0.855} & \textbf{0.092} & \textbf{0.968} & \textbf{0.884} & \textbf{0.077} & \textbf{0.968} \\\hline
\end{tabular}

 \label{tab:ablationFusion}
\end{table*}

\vspace{-2mm}
\section{Experiments}
\label{sec:Experiments}
\vspace{-1mm}
\subsection{Data collection}
All the data samples are from Chang Gung Research Database~\cite{CGRD2017}, Chang Gung Memorial Hospital, Taiwan. We follow the Helsinki declaration with ethical permission number IRB-202100564B0 (The correlation between chest x-ray and bone density). In the database, we searched patients with both DXA and CXR taken within 2 weeks from patients undergoing annual health checkups. Patient information has been removed from the data source to protect privacy. The data are original DICOM images of chest plain film. And each plain film is linked to BMD values. The DXA machine is GE lunar, X-ray detector is Canon CXDI 710C. The CXR view is PA, the voltage is 115/120 kV, and the pixel spacing is 0.16*0.16 or 0.125*0.125 (mm*mm). We exclude unsuitable cases such as implantation and bone fracture by running quality assessment preprocessing steps in~\cite{kang2021semiBMDhip}.

\vspace{-1mm}
\subsection{Experiment Setup}
\label{subsec:exp_setup}
\noindent{\bf Dataset.} We collected 13719 frontal view CXR scans, with paired DXA BMD scores (on four lumbar vertebrae L1 - L4) as ground truth. The data is provided to us from \textit{Chang Gung Memorial Hospital} after removing privacy information. All experiments use the same data split, with 11024 and 2695 patient cases for training/validation and testing, respectively. There is no patient overlapping between data splits. The model train-val/test for different lumbar vertebrae are conducted in four separate experiments, and there is no vertebra mixing among different lumbar BMD tasks even though their inputs are the same. For a particular lumbar BMD model, it is trained using 4-fold cross-validation, with train/validation ratio of 3 to 1. The ensembles of predictions from all 4-fold models on the testing set are reported as final results.



\begin{figure*}[h]
	\centering
    \hfill
	\includegraphics[width=0.42\linewidth]{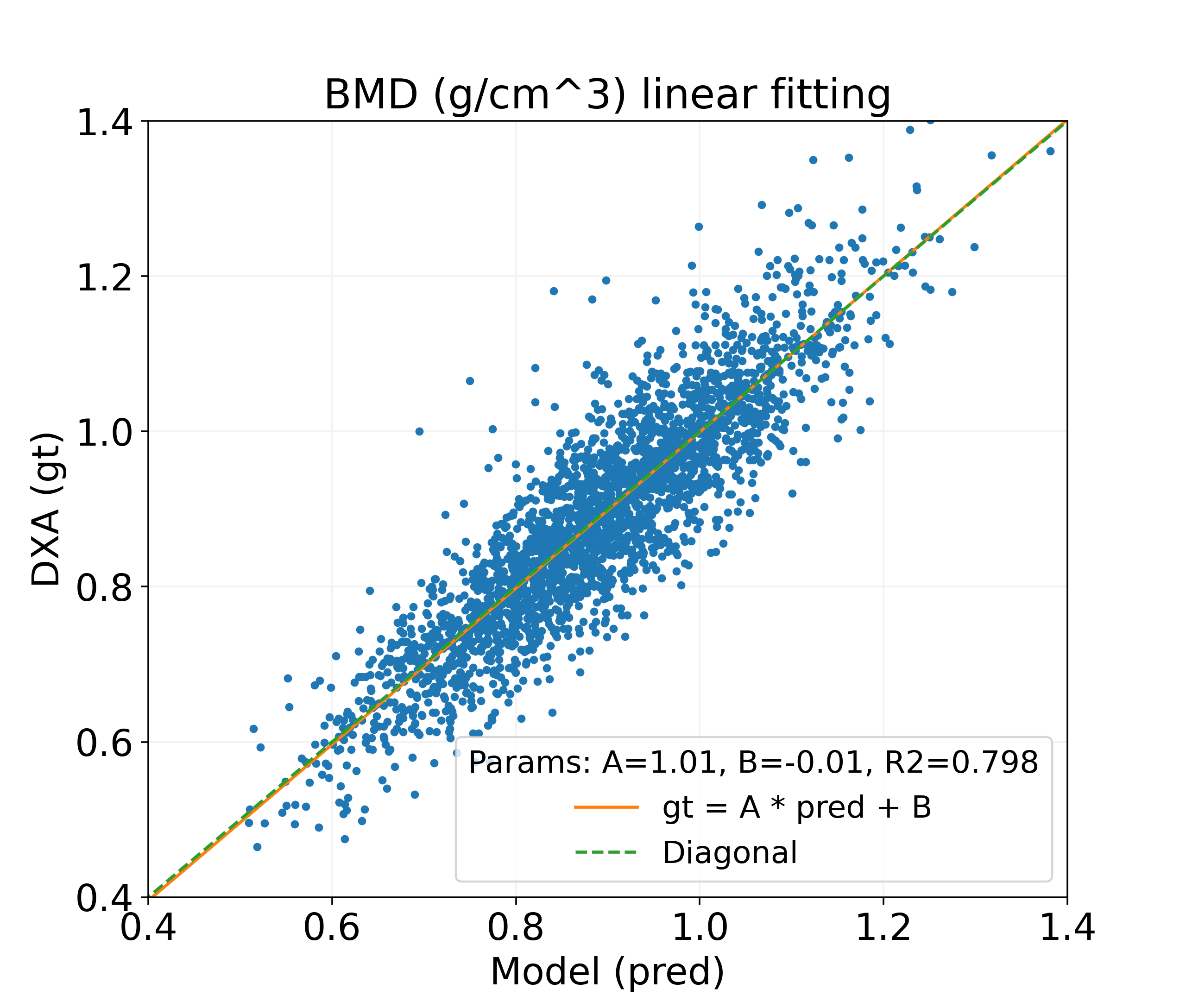}
	\hfill
	\hfill
	\includegraphics[width=0.45\linewidth]{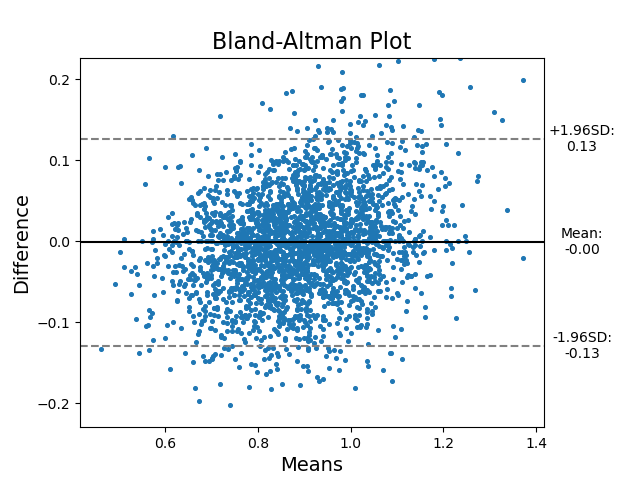}
	\hfill
	\caption{Illustrate the proposed model results on Lumbar 1 BMD, each prediction is compared with its paired DXA BMD value. In the linear fitting line (left), \textbf{A} and \textbf{B} represent \textbf{slop} and \textbf{y-intercept}, \textbf{R2} is the coefficient of determination. In the Bland-Altman plot (right), the horizontal axis is the mean, and the vertical axis is the difference between each pair.}
	\label{fig:result_attentiveMultiROI_fittingCurves}
\end{figure*}

\subsection{Data distribution}
\subsubsection{Patient statistics by gender and age}
Our data is from annual body check-ups in a regional hospital, whose distribution should be clearly stated before performance discussion. The data scope in terms of race or ethnicity is limited, and the BMD-related characteristics would not represent a general population precisely. In our task setting, we only select patients aged from 40 to 90 in the training/testing sets because this age range covers most osteoporosis onsets~\cite{who1994assessment} and other age groups do not have enough samples. In Figure~\ref{fig:data_age_bmds} (1), the data counts peak between 50 and 60. 

\subsubsection{The mean BMD values and caveats}
In Figure~\ref{fig:data_age_bmds} (2)(3), each curve describes the average BMD values at each age vertebra-wise. Data beyond 75 years old is not shown out of sample scarcity and large BMD variance. Lumbar 1 has the smallest mean value at all ages. Male and female have different BMD trends, and the female average BMD values drop sharply after the age of 50. The bumps or up-risings in the BMD curves result from the data source (annual checkups) characteristics. So the BMD distribution may not be consistent between neighboring age groups. Take female Lumbar 4 BMD values as an example, the 95\% Confidence Intervals (\textbf{CI}) of patients aged 74, 75, 76, 77, 78, 79, 80 are 0.902$\pm$0.394, 0.882$\pm$0.449, 0.912$\pm$0.494, 0.913$\pm$0.288, 0.88$\pm$0.272, 1.003$\pm$0.55, 0.944$\pm$0.406 respectively. 

\subsubsection{Confidence Intervals by BMD status}
The model only utilize BMD values during training/inference, ignoring the gender or age information. The 95\% CIs in BMD range for Lumbar 1 \textit{normal}, \textit{osteopenia}, \textit{osteoporosis} respectively are 1.0$\pm$0.17, 0.81$\pm$0.08, 0.65$\pm$0.1, for Lumbar 2 are 1.07$\pm$0.21, 0.85$\pm$0.08, 0.68$\pm$0.1, for Lumbar 3 are 1.14$\pm$0.23, 0.9$\pm$0.08, 0.74$\pm$0.11, for Lumbar 4 are  1.13$\pm$0.25, 0.88$\pm$0.08, 0.72$\pm$0.12. Notably, Lumbar 1 osteopenia (38\%) and osteoporosis (11\%) amount similarly to normal (51\%) cases, while Lumbar 4 normal amount (70\%) is much larger than osteopenia (23\%) and osteoporosis (7\%).

\begin{table*}[h]
	\caption{The Attentive Multi-ROI model classification characteristics using different prediction thresholds. Ground truth osteoporosis uses T-score -2.5 as judging threshold. Prediction classification use either unified (-1.75, -2, -2.25, -2.5) T-score thresholds for all vertebra or \textit{Flex} thresholds. \textit{Flex} Thresholds are -2.2,-2.1,-2.0,-1.9 for L1,L2,L3,L4 respectively. }
	\centering
	
\scriptsize

\begin{tabular}{|c|c|c|c|c|c|c|c|c|c|c|c|c|c|c|c|c|c|c|}
\hline
\multirow{2}{*}{\begin{tabular}[c]{@{}c@{}}T-score\\ Thresholds\end{tabular}} & \multicolumn{2}{c|}{L1} & \multicolumn{2}{c|}{L2} & \multicolumn{2}{c|}{L3} & \multicolumn{2}{c|}{L4} & \multicolumn{2}{c|}{Average} \\\cline{2-11}
& Sens       & Spec      & Sens       & Spec      & Sens       & Spec      & Sens       & Spec      & Sens         & Spec         \\\hline
-1.75                                                                         & 93.9\%     & 85.1\%    & 91.7\%     & 89.1\%    & 93.4\%     & 91.1\%    & 85.0\%     & 93.1\%    & 91.0\%       & 89.6\%       \\\hline
-2                                                                            & 86.2\%     & 91.0\%    & 85.0\%     & 93.0\%    & 84.8\%     & 94.8\%    & 75.4\%     & 95.4\%    & 82.8\%       & 93.6\%       \\\hline
-2.25                                                                         & 77.6\%     & 95.3\%    & 76.3\%     & 95.8\%    & 72.0\%     & 96.9\%    & 68.9\%     & 97.4\%    & 73.7\%       & 96.4\%       \\\hline
-2.5                                                                          & 61.9\%     & 97.5\%    & 63.5\%     & 97.9\%    & 60.7\%     & 98.3\%    & 54.5\%     & 98.9\%    & 60.2\%       & 98.1\%       \\\hline
Flex                                                                          & 79.8\%     & 94.7\%    & 82.3\%     & 94.4\%    & 84.8\%     & 94.8\%    & 78.4\%     & 94.4\%    & 81.3\%       & 94.6\% \\ \hline
\end{tabular}

 \label{tab:AttentiveMultiROISensSpec}
\end{table*}

\vspace{2mm}
\subsection{Performance Metrics}
We evaluate our models on quantitative metrics essential to clinical verification. The vertebra-level metrics include the \textit{Root-Mean-Square Error} (RMSE), the \textit{Pearson Correlation Coefficient} (R-value), \textit{Area Under Curve} (AUC), sensitivity, specificity, coefficient of determination (R squared or $R^2$) of the linear fitting curve, standard deviation of the prediction errors. The patient-level metrics are sensitivity and specificity. RMSE measures the root of averaged square differences between the predicted and ground truth. The R-value measures the linear correlation between the predicted and the ground truth, only considering the sequential correlation regardless of the absolute values. For osteoporosis classification, BMD values are transformed into T-score values by checking the transforming table in the DXA machine~\cite{who1994assessment}\cite{Compston2017UKCG}. In the T-score range, the AUC measures accumulated true positive (osteoporosis) rate under different judging thresholds for osteoporosis classification. The sensitivity and specificity are also for classification purposes. The linear fitting curve illustrates the general correspondence between prediction and ground truth. The coefficient of determinant quantifies the fitting goodness. We also draw the Bland-Altman plot which shows the standard deviation limits and prediction error distribution.

\subsection{Attentive Multi-ROI model performance (vertebra level) }
\label{ssec:implement_performance}
\begin{figure}[bt!]
	\centering
	\includegraphics[width=0.97\linewidth]{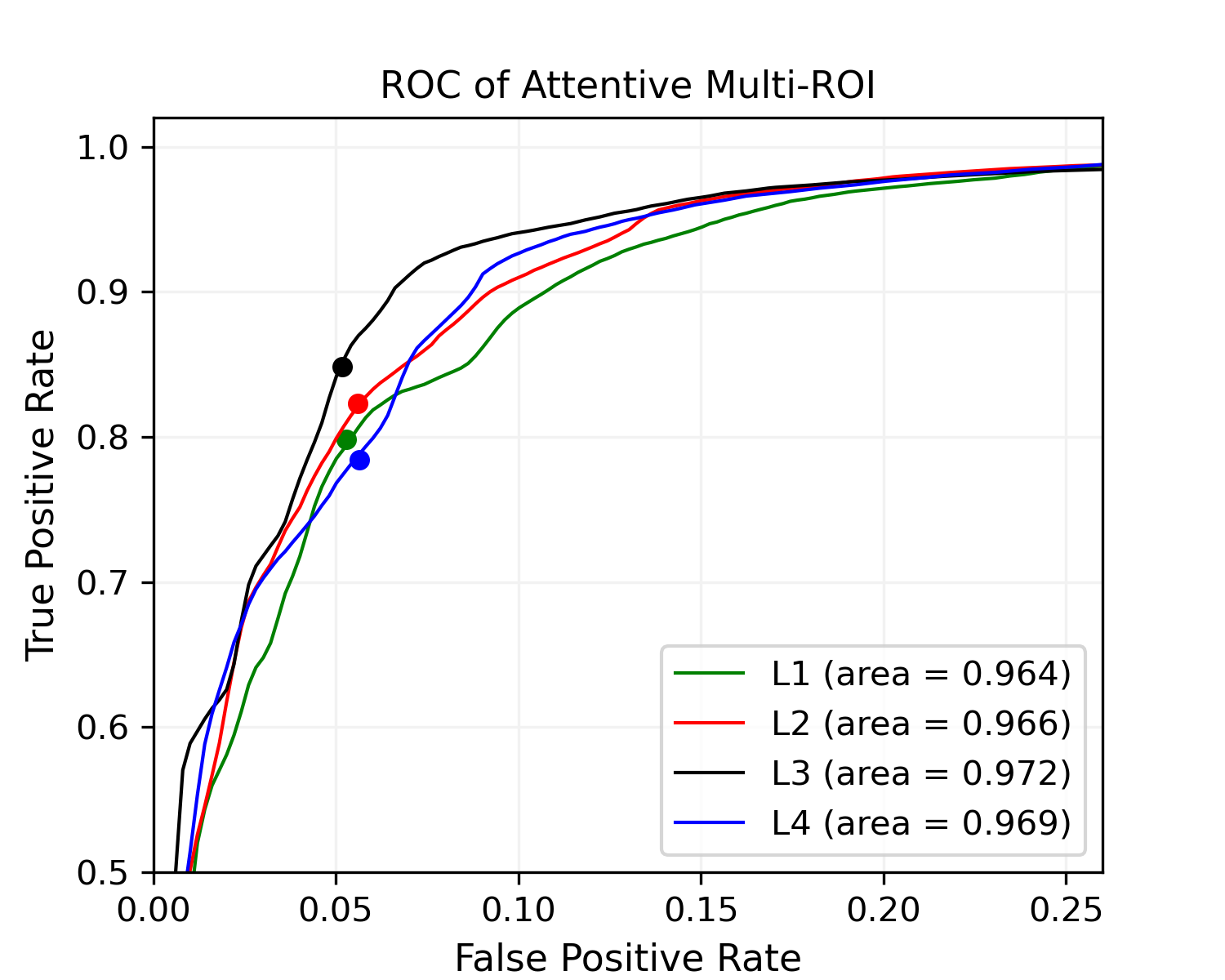}
	\caption{The Receiver Operating Characteristic (ROC) Area Under Curve (AUC) of the proposed \textit{Attentive Multi-ROI} model for osteoporosis classification in L1, L2, L3, L4 experiments (only upper left AUC). The colored dots operate on \textbf{Flex} T-score thresholds.}
	\label{fig:attentiveMultiROIROC}
\end{figure}

Each lumbar BMD model is trained four times using a 4-fold cross-validation setting. The prediction ensemble of these models on testing set is recorded as the final result. Due to the model variants, there are 5 records for each lumbar BMD task in Table~\ref{tab:ablationFusion}. On all task metrics, the Attentive Multi-ROI model achieves the best performance except L2 AUC. The proposed model outperforms others by approximately 1\% in terms of R-value for all BMD tasks, which clearly demonstrates its superiority. The proposed also has pretty low RMSE values, especially on L1 and L2, supporting the prediction accuracy. While the R-value and RMSE are direct error measures, AUC scores evaluate osteoporosis classification statistically in Figure~\ref{fig:attentiveMultiROIROC}. L4 has a larger AUC score because its osteoporosis ratio (7\%) is much smaller than normal (70\%) or osteopenia (23\%), while L1 is the opposite. We show the sensitivity and specificity in Table~\ref{tab:AttentiveMultiROISensSpec}. Applying the \textbf{Flex} thresholds, the model achieves high averaged sensitivity (81.3\%) and specificity (94.6\%).

To show the performance intuitively, we draw the linear fitting line and the Bland-Altman plot for L1 predictions in Figure~\ref{fig:result_attentiveMultiROI_fittingCurves}. The \textit{intersect} (-0.01, close to 0) and \textit{slop} (1.01, close to 1) of the linear fitting line demonstrates the general correctness, and the \textit{R-squared} (0.798) measures the closeness between predictions and the ground truth. In the Bland-Altman plot, value errors are drawn against value means for each prediction and DXA BMD pair. The relatively small standard deviation (0.065) and the concentrated scattering further prove the performance consistency. Outliers beyond the $\pm$1.96SD limits occupy a small portion of all. Linear fitting and Bland-Altman plots of other lumbar vertebrae lead to similar conclusions. Error analysis is in the Ablation section.

\subsection{The patient-level osteoporosis classification}
\begin{figure}[bt!]
	\centering
	\includegraphics[width=0.97\linewidth]{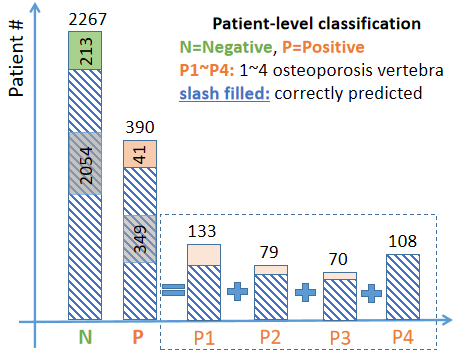}

	\caption{Patients with four lumbar records (2657 totally) in the testing set are assigned as normal (Negative) or osteoporosis (Positive). The positive cohort can be further decomposed into four bins, according to the number of osteoporosis vertebrae. The shadowing parts are true negatives and true positives from the Attentive Multi-ROI model, applying $Flex$ thresholds.}
	\label{fig:patientLevelClassification}
\end{figure}

For opportunistic alarming, the goal is to inform osteoporosis risks. Although we predict the BMD of four lumbar vertebrae with four separate models, we aim to generate unified alarming signals. Therefore we calculate the patient-level osteoporosis classification performance. Each patient is either \textbf{normal} (all four lumbar T-scores are larger than $-2.5$) or \textbf{osteoporosis} (any lumbar vertebra has a T-score smaller than $-2.5$). The patient distribution is in Figure~\ref{fig:patientLevelClassification}, where there are 2267 normal cases, 390 osteoporosis cases. 

Applying different T-score thresholds, we calculate the sensitivity and specificity to evaluate the proposed model in Table~\ref{tab:patientLevelClass}. Referring to the proper vertebra-level thresholds in Table~\ref{tab:AttentiveMultiROISensSpec}, we only compare four thresholds for the patient-level classification. These settings cover both high sensitivity ($0.933$) and high specificity ($0.966$) for osteoporosis. As a balanced configuration, the $Flex$ thresholds (T-score $-2.2$,$-2.1$,$-2.0$,$-1.9$ for Lumbar 1,2,3,4 respectively) achieves approximately $90\%$ for both osteoporosis sensitivity and specificity. It also achieves $99.7\%$ patient-level osteopenia specificity, implying picking out nearly no patients with healthy BMD. These results strongly support the practical applicability of our proposed model. Applying the $Flex$ thresholds in Figure~\ref{fig:patientLevelClassification}, the model can pick out osteoporosis patients very well, with $75.2\%$ \textbf{P1}, $92.4\%$ \textbf{P2}, $97.1\%$ \textbf{P3}, $100\%$ \textbf{P4}.

\subsection{The model variants}
\subsubsection{The Baseline model}
\label{ssec:implement_baseline}
There is no shortage of established CNN architectures~\cite{resnet}\cite{vgg}, and we can simply employ off-the-shelf CNNs as the \textit{Baseline} model. Since the VGG and Resnet have been shown to work well on hip X-ray BMD estimation\cite{kang2021semiBMDhip}, they may also succeed in the CXR-based BMD estimation. In the \textit{Baseline} model, we adopt VGG16 as the feature extractor, apply the Global Average Pooling (GAP) to reduce spatial dimension, use two linear layers with \textit{ReLu} non-linearity in the regressor to predict BMD, and use Mean Squared Error (MSE) loss to train the model. Among different \textit{Baseline} input candidate ROIs, the whole chest performs the best. Modalities such as lumbar ROI or cervical ROI get 2\% to 5\% lower R-value than the whole chest ROI. In Table~\ref{tab:ablationFusion}, the \textit{Baseline} model serves as a reference for performance comparison.

\subsubsection{The Multi-ROI model (MulROI) } 
\label{ssec:implement_multiROI}
In order to investigate the effect of the \textit{Transformer Encoder} in the proposed Attentive Multi-ROI model, we replace \textit{attentive feature fusion} with \textit{direct feature concatenation} in Figure~\ref{fig:plain_fusion} in the \textit{Multi-ROI} model. The concatenated global feature has the length of \textit{512*15}, and the global regressor now has a larger input dimension. The proposed ( Attentive Multi-ROI) model has consistent advantages over the \textit{MulRoi} (plain-fusion Multi-ROI) in Table~\ref{tab:ablationFusion}, which demonstrates the positive effect of \textit{Transformer Encoder}.

\subsubsection{The Attentive Multi-Patch model (AttMulPat)}
\label{ssec:implement_AttMultiPatch} 
To investigate the benefits of precise ROI extraction in the proposed \textit{Attentive Multi-ROI}, we replace the \textit{landmark-based ROI extraction} with the \textit{image patch splitting} in the \textit{Attentive Multi-Patch} model (AttMulPat). We split the high-resolution chest X-ray image into evenly distributed patches in Figure~\ref{fig:image_patches}. Though lacking the precise landmark-based cropping, the \textit{AttMulPat} model is able to learn both individual patch details and inter-patch relations. However, the representative meaning of each patch is less certain due to the scanning variations in patient postures and body sizes. From the comparison of the \textit{AttMulPat} and \textbf{Proposed} in Table~\ref{tab:ablationFusion}, the landmark detection and precise ROIs benefit all four lumbar BMD tasks.

\subsubsection{The Multi-Patch model (MulPat)}
\label{ssec:implement_MultiPatch}
To show the effect of \textit{Transformer Encoder} on \textit{Attentive Multi-Patch} model, we train and test the \textit{Multi-Patch} model which instead uses the \textit{plain concatenation}. Their comparisons in Table~\ref{tab:ablationFusion} again show that the attention module consistently benefits all four tasks with better feature fusion ability. 



\subsection{Performance comparisons}
\label{ssec:implement_comparison}
To see the advantage of the \textit{Multiple-Modality} inputs working flow in our four models (\textit{MulPat, AttMulPat, MulRoi, AttMulRoi,}), we compare them with the \textit{Baseline} in Table~\ref{tab:ablationFusion}. The \textit{Multiple-Modality} models could extract not only global patterns but also detailed local textures, thus having much better results. With simple patch splitting and attentive fusion, the \textit{AttMulPat} model outperforms the \textit{Baseline} model significantly in terms of R-value and RMSE. 

The four \textit{Multiple-Modality} models differ by the \textit{ROI extraction} component and the \textit{global fusion} component. To see the component-wise boosting effects from the \textit{landmark-based ROI extraction} and the \textit{Transformer Encoder}, four models are compared to each other with the base being the \textit{MultPat} model. Applying the \textit{Transformer Encoder} only (AttMulPat) or applying the \textit{landmark-based ROI extraction} only (MulRoi) contributes similar amount of benefit (about 1\% R-value boosting) to the \textit{MultPat} model in Table~\ref{tab:ablationFusion}. Applying these two simultaneously in our proposed \textit{AttMulRoi} model leads to the best performance, with 2\% R-value boosting. On the one hand, the precise ROI croppings in the \textit{MulRoi} and \textit{AttMulRoi} models enable more efficient local texture utilization. On the other hand, the plain concatenation in the \textit{MultPat} and \textit{MulRoi} models treat all the individual ROI features as equal which renders the model less robust to occlusion or noises, especially in case of implants or tissue consolidations. The \textit{Transformer Encoder} adjusts the individual features in a learnable and flexible manner, addressing the correlations among chest bones, which leads to improved feature robustness.

\begin{table}[h]
	\caption{Patient-level sensitivity and specificity. Unified thresholds (-1.75, -2, -2.25) ignore the lumbar BMD differences, while $Flex$ (thresholds -2.2,-2.1,-2.0,-1.9 for Lumbar 1,2,3,4) is aware.}
	\centering
	\scriptsize

\begin{tabular}{|c|c|c|c|c|}
\hline
\multirow{2}{*}{\begin{tabular}[c]{@{}c@{}}T-score\\ Thresholds\end{tabular}} & \multicolumn{2}{c|}{Osteoporosis} & \multicolumn{2}{c|}{Osteopenia} \\\cline{2-5}
& Sens            & Spec           & Sens           & Spec          \\\hline
-1.75                                                                         & 0.933           & 0.873          & 0.453          & 0.995         \\\hline
-2                                                                            & 0.844           & 0.929          & 0.341          & 0.998         \\\hline
-2.25                                                                         & 0.736           & 0.966          & 0.255          & 0.999         \\\hline
Flex                                                                          & 0.895           & 0.906          & 0.392          & 0.997        \\\hline
\end{tabular}


 \label{tab:patientLevelClass}
\end{table}

\begin{figure*}[bt!]
	\centering
	\includegraphics[width=0.24\linewidth]{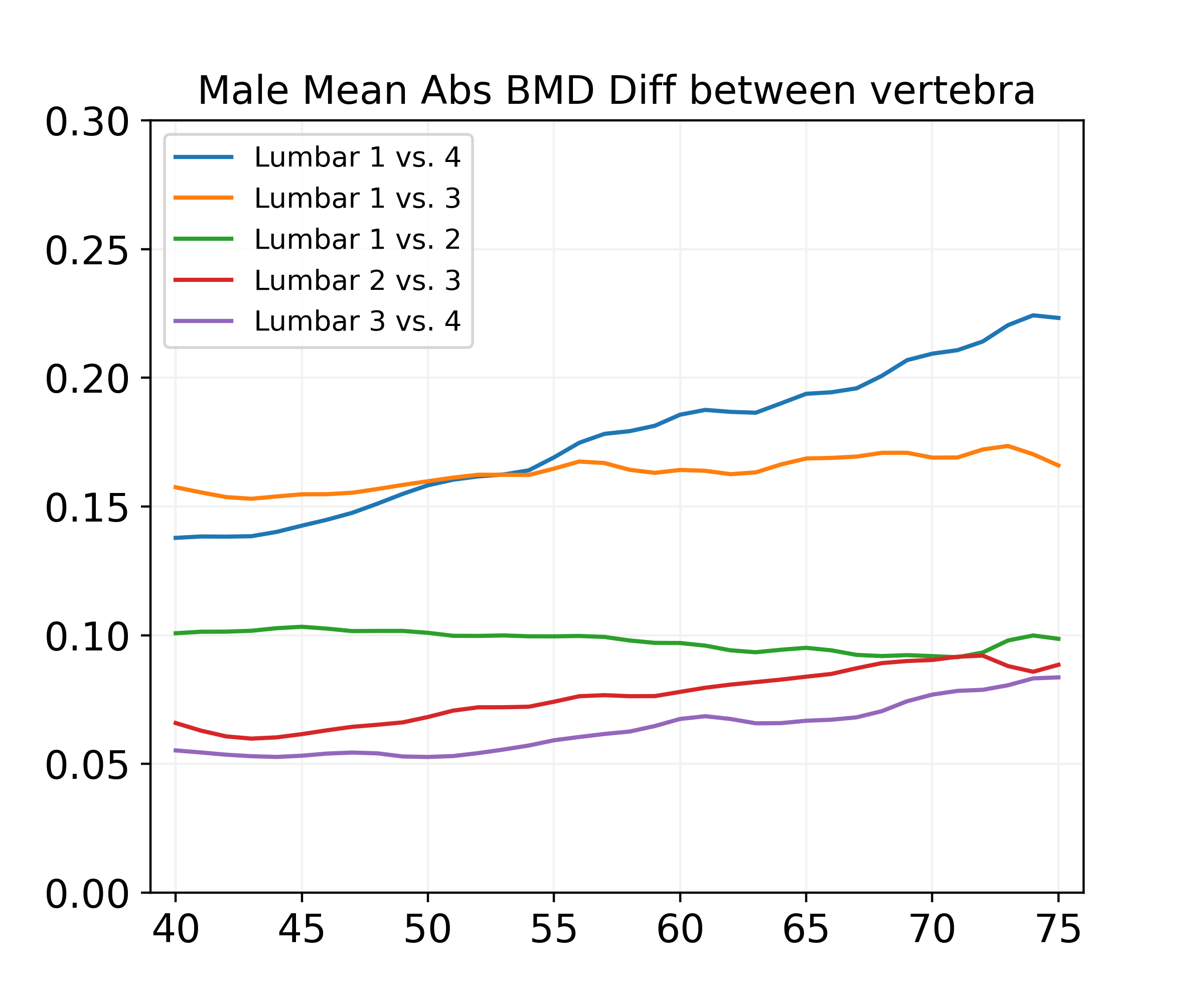}
	\includegraphics[width=0.24\linewidth]{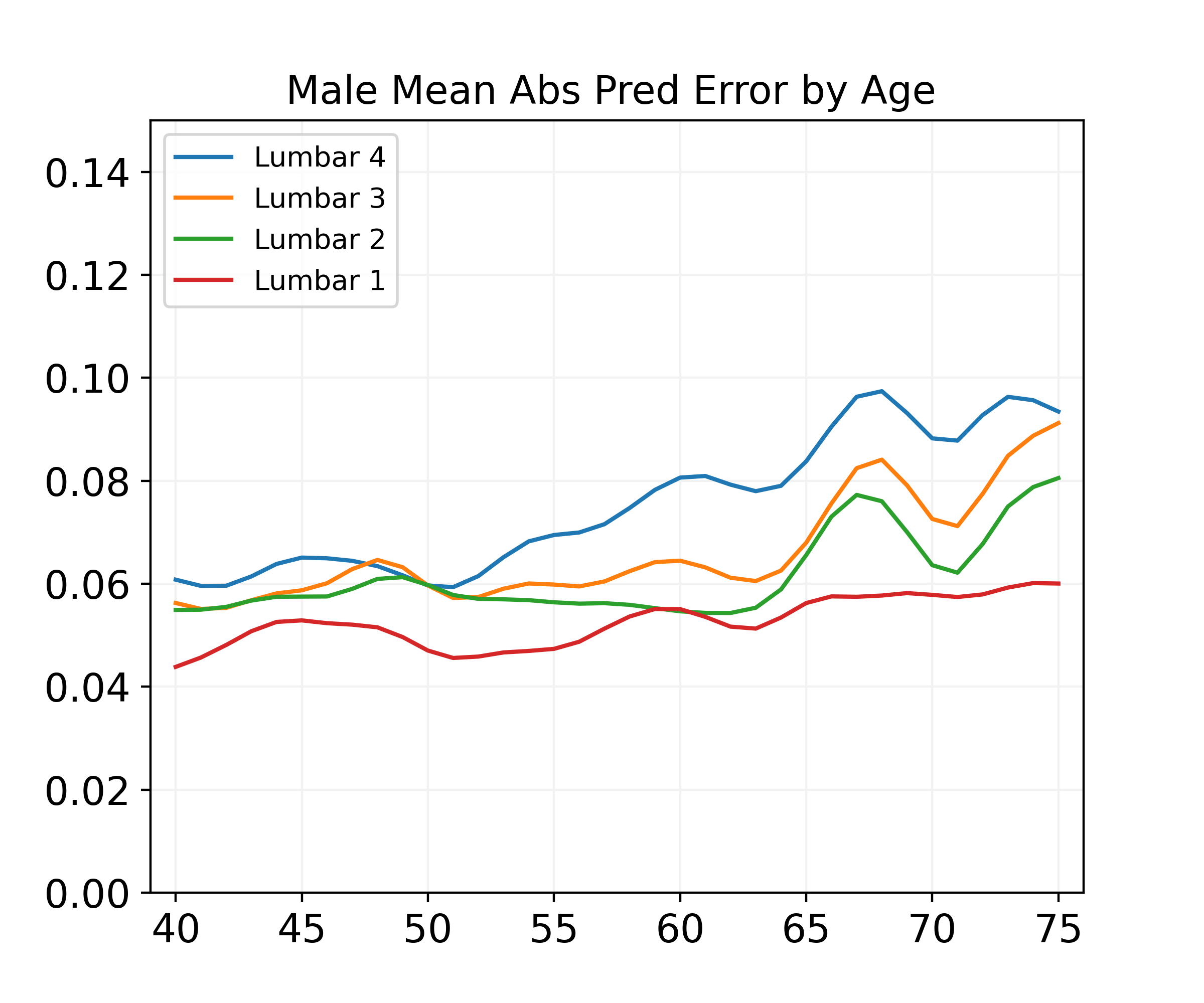}
	\includegraphics[width=0.24\linewidth]{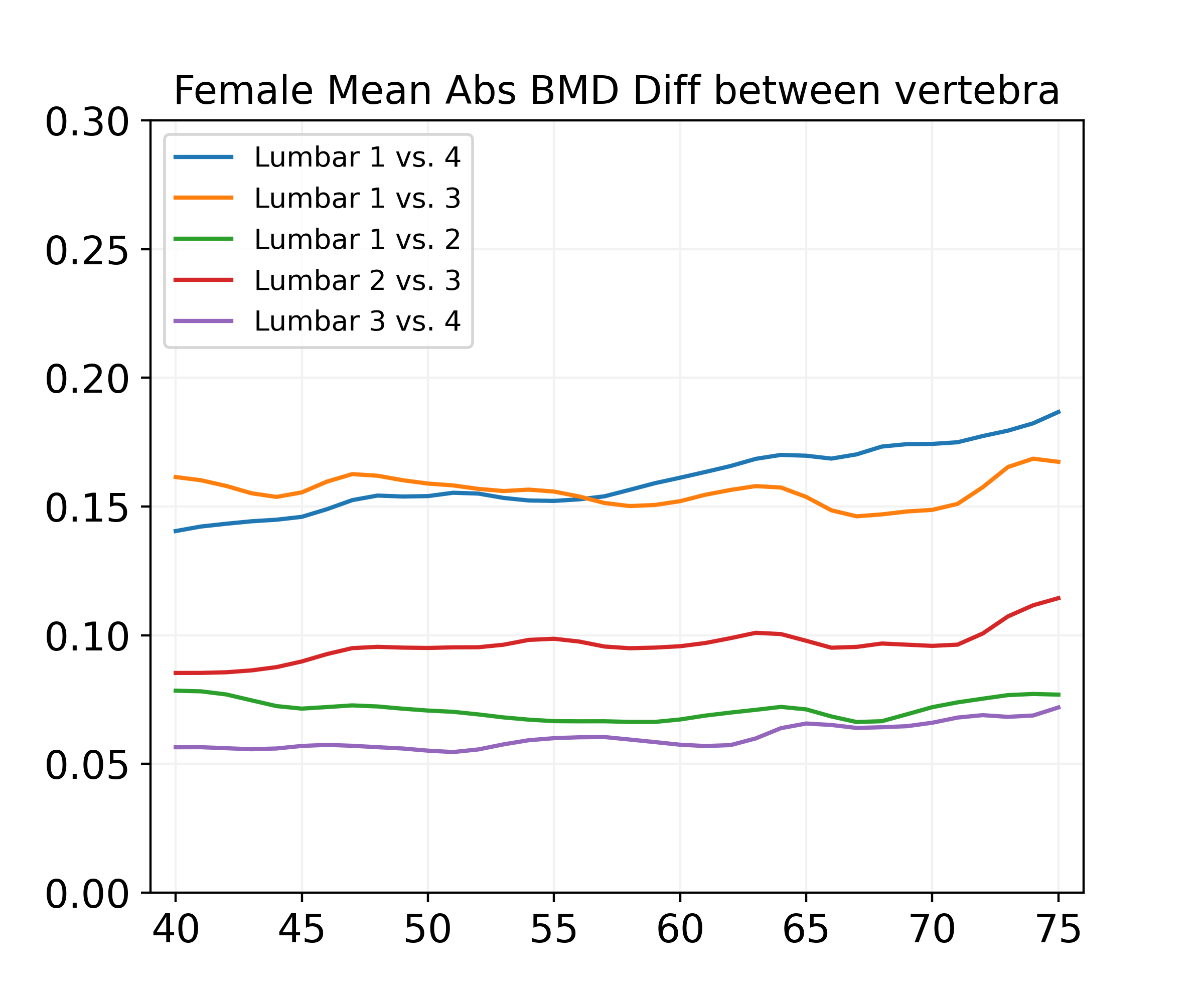}
	\includegraphics[width=0.24\linewidth]{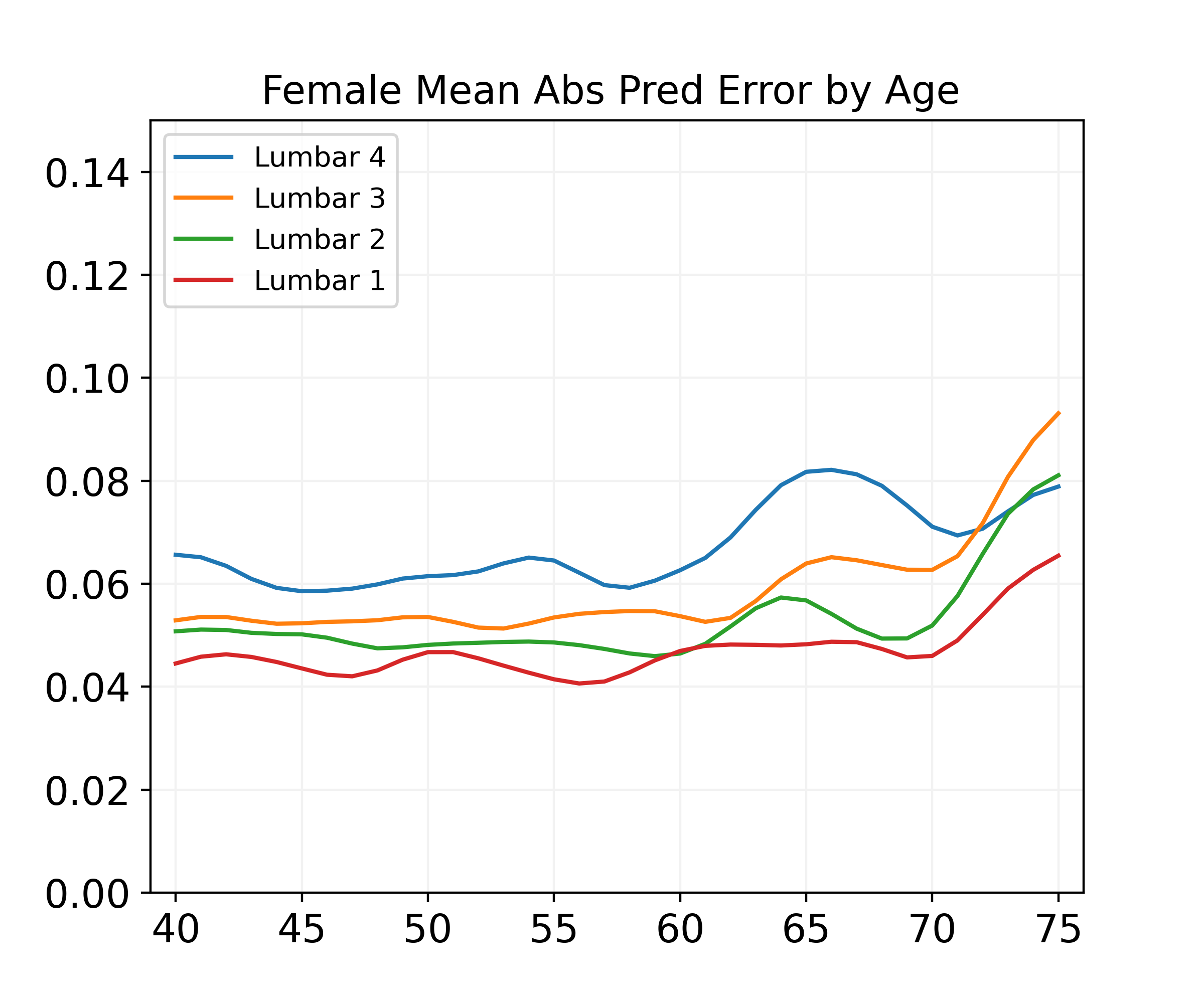}
	\caption{ Cross-vertebra BMD differences (1)(3). Measure the mean of absolute difference between vertebra pairs, using DXA BMD ($g/cm^3$). The mean of absolute prediction error (2)(4). The proposed model (Attentive Multi-ROI) predictions have satisfactory error range. }
	\label{fig:data_absDiff}
\end{figure*}

\section{Ablation study}
\subsection{Convolutional backbone selection}
To find the best convolutional backbones for feature extraction, we compare VGG13, VGG16, VGG19, Resnet18, Resnet34, Resnet50~\cite{Simonyan15VGG}\cite{resnet} with the \textit{Baseline} working flow. The input modalities have a fixed resolution of \textit{(256, 256)}. In our experiments, VGG16 and VGG19 outperform other backbones in different lumbar BMD prediction tasks, in terms of R-value, RMSE, AUC. The VGG family in general surpasses Resnet variants in all tasks, suggesting that relatively simple backbones work better on CXR texture recognition. The proper combination of convolutional kernels and backbone depth in VGG16 exploits the CXR textures better than VGG19, and we use VGG16 as the default feature extractor in all experiments.

\subsection{Image splitting dimension for the Multi-Patch model}
In the \textit{Multi-Patch} model, we split the original CXR image into even patches in Figure~\ref{fig:image_patches}. We train and test the \textit{Multi-Patch} model with different split dimension \textit{N} (\textit{N} rows by \textit{N} columns, \textit{N}=2,3,4). Smaller split dimensions (\eg \textit{N} = 2) produce patches covering a larger field of view, which may impede finer-level texture exploration. Larger split dimensions (\eg \textit{N} = 4) produce patches covering a smaller area, leading to theme shifting in individual patches because of the varied shape, size, and positions during scanning. In the comparison, \textit{N}=3 has the best performance (used as default).

\subsection{Determine the proper T-score thresholds}
\label{thresholdapplicability}
Different lumbar vertebrae have distinct BMD distribution in Figure~\ref{fig:data_age_bmds}. We have known their varied Confidence Intervals for osteoporosis, osteopenia, and normal cases in previous sections. There are many cases whose lumbar 4 is normal/osteopenia but the lumbar 1 could be osteoporosis using the same T-score threshold (-2.5). In the opportunistic setting, the models not only predict individual vertebra BMD, but also aim to generate unified alarming information. Therefore the T-score thresholds for model predictions on different vertebrae may be adjusted to get consistent sensitivity and specificity. 

In Table~\ref{tab:AttentiveMultiROISensSpec}, we investigate osteoporosis classification under different T-score thresholds. To get a balanced and applicable result in practice, the sensitivity may be desired above 80\%, specificity to be above 90\%. Under the fixed T-score threshold (-1.75, -2, -2.25 or -2.5), the sensitivity and specificity have different ratios across vertebrae. When it is reasonably good on Lumbar 1, the metrics are tilted towards a low sensitivity on Lumbar 4. A balanced performance means similar metrics on all vertebrae, which requires flexible threshold adjustments. The \textbf{Flex} achieves this goal with approximately 80\% sensitivity and 94\% specificity on all vertebrae. The \textbf{Flex} thresholds produce an equalized alarming degree across vertebrae, helpful for unified patient-level judgments.

\subsection{Factors leading to large prediction error}
\label{ssec:postProcess}
There can be many factors influencing predictions, such as the hardware, scanning settings, and the sample characteristics ( gender, age, lumbar vertebra ). For results on Lumbar 1 in Figure~\ref{fig:result_attentiveMultiROI_fittingCurves} (1), points far away from the \textit{Diagonal} are large error predictions, corresponding to points above +1.96SD or below -1.96SD in (2). Beyond [-1.96SD, +1.96SD] range in L1 Bland-Altman plot, there are 144 cases. Among them are 68 female and 76 male, which are not gender specific. They scatter across all ages, with data count distribution similar to Figure~\ref{fig:data_age_bmds} (1). We also check their scanning hardware settings such as voltage, image spacing, image size, without finding any particular insights. Then we inspect their landmark localization and bone patch extraction procedures, but there are no mistakes in the intermediate results either. Visually checking these cases, there is no differences from the more accurately predicted. So the large errors are due to unknown factors, remained for future research.

\subsection{The model performance boundary}
In our task setting, the lumbar BMD is estimated from the chest radiology scanning. Due to the physiological differences between body parts and bones, the BMD of these areas would have statistical variances. Therefore the cross-bone BMD variances could provide a good hint on the model performance boundary. To get the cross-bone BMD discrepancy, we calculate the mean absolute DXA BMD differences between vertebrae, plotted by gender and age in Figure~\ref{fig:data_absDiff} (1)(3). The neighboring bones have relatively small and stable differences, such as Lumbar 1 versus 2, and Lumbar 2 versus 3. As the distance increases, the cross-bone BMD difference increases, such as Lumbar 1 versus 3, and Lumbar 1 versus 4. The average BMD differences between two neighboring bones stay in $[0.05, 0.10]$ range for the most time.

To examine the model performance, we plot the Mean Absolute Error (MAE) between prediction and DXA BMD by gender and age in Figure~\ref{fig:data_absDiff} (2)(4). Generally, predictions on Lumbar 1 have a smaller absolute error, while Lumbar 4 predictions have a larger error, which is in accordance with their BMD magnitude distribution. For most cases, the prediction errors fall in $[0.04, 0.08]$ range, which is even smaller than the counterparts of neighboring bones. Referencing the mean absolute difference in (1)(3), it serves as the upper bound for cross-bone BMD estimation which implies the model performance boundaries. By comparing (1)(2) or (3)(4) in Figure~\ref{fig:data_absDiff}, our proposed model marches near this upper limit.

Besides MAE, the DXA BMD R-values between vertebra pairs can also be used to explore the performance boundary. The DXA BMD R-value between L1 and L2 is 0.918, between L1 and L3 is 0.878, and between L1 and L4 is 0.807. The model prediction R-values are 0.894, 0.899, 0.887 on L1,L2,L3 respectively in Table~\ref{tab:ablationFusion}. Though neighboring bones (L1 and L2) have a higher BMD correlation than CXR-based prediction, the model predictions have surpassed the unconnected bones (L1 and L3) in providing BMD reference. Given the closeness between vertebrae in terms of both geometric distance and physiological function, our CXR-based model has achieved remarkable performance.

\section{Discussion}
\subsection{The ground truth DXA BMD limitations}
The DXA scan is a 2D projection of the 3D object, unavoidably including noise from posterior parts of the vertebrae to interfere with the vertebral body BMD~\cite{3DshapeBMDFromDXAscan}\cite{shapeDensityThicknessfromDXA}. Prior works extracted 3D geometric and structural measurements from area-DXA scans to mitigate this shortcoming~\cite{3DshapeBMDFromDXAscan}\cite{shapeDensityThicknessfromDXA}\cite{extract3DinforViaDXA}\cite{3DDXA}. Metrics such as the trabecular bone score (TBS, based on lumbar DXA scanning) are developed to provide bone microarchitecture and skeletal information~\cite{Silva2014TrabecularBS}\cite{Bousson2011TrabecularBS}\cite{Harvey2015TrabecularBS}. These DXA augmentations shine insights for accuracy correction and quality assessment. Though the opportunistic applications do not require strict accuracy, we should be aware of the limitations of using DXA BMD as the ground truth.


\subsection{Data source limitations}
The data is NOT a randomly collected sample set. Instead, it is a convenience sampling from 'annual health checkup' population (Taiwan) who need to pay their fees. They are in general healthy enough not to visit clinics. Therefore, the sampling may not reflect the general trends of declining BMD. People younger than 40 or older than 90 are not included due to sample scarcity. In this study, the patient statistics with respect to gender and age may not accurately reflect clinical reality. We exclude the cases with implants or bone fractures, which are less frequent in the opportunistic setting. Some characteristics and limitations have been discussed in the data distribution section. As our data is from one hospital, the model must be tested in more centers with different hardware scanners before wide-range clinical application.

\subsection{Result interpretation limitations}
Although deep learning based models have been successfully applied in many vision and language tasks, application in medical tasks requires more caution. The BMD-related patterns and texture are not visually identifiable by a human. In our experiments, the correlation between chest X-ray image and lumbar DXA BMD is established through training models to predict the paired information. The functioning principles of this process have not been fully examined. Although our model achieves impressive osteoporosis sensitivity and specificity, a finer-level analysis of BMD correlation and prediction is expected in future studies. 

\subsection{Applicability}
The performance of using chest x-ray to predict BMD is unlikely to match direct DXA examination on hip and lumbar. Part of this is explained in the performance boundary section, where L1 and L2 have a higher R-value than model predictions. For the formal judgment of osteoporosis or osteopenia, only DXA BMD on the lumbar or hip should be considered~\cite{who1994assessment}\cite{Compston2017UKCG}. CXR-based BMD prediction works as a low-cost and opportunistic way to make alarms instead determining osteoporosis status, and the patient should take hip/lumbar DXA scans in cases of positive predictions. For people older than 70 or with spine diseases, hip BMD is more accurate. Certain population such as post-menopause females should take comprehensive examinations guided by medical experts~\cite{Compston2017UKCG}. 




\section{Conclusion}
In this paper, we design deep learning models to estimate lumbar vertebra BMD from chest X-ray images. We propose the anatomy-aware \textit{Attentive Multi-ROI} model that can extract local bone textures and generate robust feature representation. The landmark-based ROI extraction promotes the local feature reliability against scanning variations. The transformer-based encoder improves the system robustness in case of noises, and occlusions. The proposed model achieves good performance on vertebra-level BMD prediction as well as patient-level osteoporosis classification. We conduct detailed comparisons of data distribution and model performance. Through extensive experiments and comprehensive analysis, the model holds great clinical potential for opportunistic screening.


%




\ifCLASSOPTIONcaptionsoff
  \newpage
\fi





%



\bibliographystyle{IEEEtran}
\bibliography{mybibliography}

%











\end{document}